\documentclass[iop,numberedappendix]{emulateapj}

\usepackage{natbib}
\usepackage{graphicx}
\usepackage{amsmath}
\usepackage{amssymb}
\usepackage{amsfonts}
\usepackage{bm}
\usepackage{epsf}
\usepackage{pifont}
\usepackage{tablefootnote}
\usepackage[perpage]{footmisc}
\usepackage[breaklinks,colorlinks,citecolor=blue]{hyperref}        

\newcommand{\cmark}{\checkmark}

\def\myfig#1{./#1}

\newcommand{\kpc}{\mbox{ kpc}}
\newcommand{\Mpc}{\mbox{ Mpc}}
\newcommand{\cm}{\mbox{ cm}}
\newcommand{\erg}{\mbox{ erg}}
\newcommand{\se}{\mbox{ s}}
\newcommand{\km}{\mbox{ km}}
\newcommand{\keV}{\mbox{ keV}}

\newcommand{\grad}{\bm{\nabla}}
\newcommand{\vect}[1]{\mathbf{#1}}
\newcommand{\ie}{\emph{i.e.,} }
\newcommand{\eg}{\emph{e.g.,} }
\newcommand{\tcool}{t_{\mbox{\tiny cool}}}
\newcommand{\tff}{t_{\mbox{\tiny ff}}}
\newcommand{\fb}{f_{\mbox{\tiny g}}}
\newcommand{\kB}{k_{\mbox{\tiny B}}}
\newcommand{\Knorm}{K_{\mbox{\tiny norm}}}

\newcommand{\till}{{\mbox{--}}}

\newcommand{\myv}{\mbox{v}}
\newcommand{\myw}{\mbox{w}}

\newcommand{\mypvs}{{\lambda^s_{\mbox{\scriptsize{v}}}}}
\newcommand{\mypvf}{{\lambda^f_{\mbox{\scriptsize{v}}}}}
\newcommand{\mypw}{{\lambda_{\mbox{\scriptsize{w}}}}}
\newcommand{\mypwf}{{\lambda^f_{\mbox{\scriptsize{w}}}}}
\newcommand{\mypws}{{\lambda^s_{\mbox{\scriptsize{w}}}}}
\newcommand{\SB}{F_{\mbox{x}} }
\newcommand{\myzeta}{\zeta}
\newcommand{\ibid}{ibid}
\newcommand{\rhotot}{\rho_{\mbox{\tiny tot}}}
\newcommand{\myR}{R}
\newcommand{\unit}[1]{\bm{\hat{#1}}}
\newcommand{\constant}{\mbox{constant}}

\defcitealias{Pratt05}{P05}
\defcitealias{Piffaretti05}{P05b}
\defcitealias{Mahdavi05}{M05}
\defcitealias{Pratt06}{P06}
\defcitealias{Donahue06}{D06}
\defcitealias{Sanderson09}{S09}
\defcitealias{Johnson09}{J09}
\defcitealias{Cavagnolo09}{C09}
\defcitealias{Pratt10}{P10}
\defcitealias{Panagoulia14}{P14}
\defcitealias{Keshet12}{K12}
\defcitealias{Voit05}{V05}

\begin{document}

\title{A Dynamically Driven, Universal Thermal Profile of Galaxy Groups and Clusters}

\author{Ido Reiss\altaffilmark{1,2} and Uri Keshet\altaffilmark{1}}

\altaffiltext{1}{Physics Department, Ben-Gurion University of the Negev, POB 653, Be'er-Sheva 84105, Israel}
\altaffiltext{2}{Physics Department, Nuclear Research Center Negev, POB 9001, Be'er-Sheva 84190, Israel}


\begin{abstract}
Large scale structures such as groups and clusters of galaxies show a universal, nearly linear entropy radial profile $K(r)$. Using 16 deprojected clusters and 12 deprojected groups from the literature, we find that $K\propto r^{0.96\pm0.01}$, consistent with the mean power-law index $(0.9\mbox{--}1.1)$ of previous studies.
A similarly good fit is given by a $\tau\propto r^{0.72\pm0.01}$ ratio between cooling and free-fall times.
Both profiles slightly flatten at small radii, as $\tau$ becomes of order unity.
The entropy profile is usually attributed to self-similar shock accretion (shown to be inconsistent with the data), to non-standard heat conduction, or to turbulent heating. We argue that a dynamical mechanism is needed to sustain such a universal profile, oblivious to the temperature peak at the edge of the core and to the virial shock at the outskirts, and robust to the presence of ongoing cooling, merger, and active galactic nucleus activity. In particular, we show that such a profile can be naturally obtained in a spiral flow, which is likely to underlie most galaxy aggregates according to the ubiquitous spiral patterns and cold fronts observed.
Generalizing a two-phase spiral flow model out to the virial radius surprisingly reproduces the thermal profile.
A generalized Schwarzschild criterion indicates that observed spiral patterns must involve a convective layer, which may regulate the thermal profile.
\end{abstract}


\maketitle

\section{Introduction}

Most groups and clusters of galaxies (galaxy aggregates; henceforth GAs) show a central dense, cool core (CC) in which the radiative cooling time of the intergalactic medium (\ie intragroup or intracluster; henceforth IGM) is shorter than the age of the GA. Some quasi-steady, smoothly distributed heat injection mechanism is needed in order to balance the observed cooling, and to sustain the only mild (factor of a few, typically) temperature drop toward the center. For reviews of such CC GAs and possible heating mechanisms, see \cite{Peterson06, McNamara07, Soker10, Fabian12, Kravtsov12, Cavaliere13, Vikhlinin14}.

Energetically, core cooling can plausibly be suppressed, for example, by thermal conduction \citep{Zakamska03} that is modified by heat buoyancy instabilities \citep{Quataert08}, or by the energy output of the active galactic nucleus (AGN) in the central cD galaxy \citep[\eg][]{Birzan04}. In particular, sufficient mechanical energy is thought to be deposited in hot AGN bubbles \citep{Churazov02}, which are found in at least $70\%$ of the CCs \citep{Dunn06}. However, it is unclear if the energy can be transferred steadily and homogeneously throughout the cooling plasma, in order to stem the local thermal instability. Moreover, a feedback mechanism is needed in order to regulate the heating and adapt to changes in the core. If the AGN output is regulated by the accreted, cooling plasma, it could furnish such a feedback loop and quench the global thermal instability as well \citep[\eg][]{Rafferty08}.

The thermal structure of an observed GA is often characterized by the azimuthally averaged, radial profiles of electron number density $n_e$ and temperature $T$. To gauge the non-adiabatic processes, in particular radiative cooling, shock heating, and heat conduction, it is useful to study the adiabat, conventionally referred to as the 'specific entropy' or just 'entropy' (henceforth), $K=n_e^{-2/3}\kB T$, where $\kB$ is the Boltzmann constant \cite[\eg][]{Voit05,Cavagnolo09}.

Using a simplified spherical model, and in the absence of radiative cooling, AGN feedback, and star formation, \cite{Tozzi01} found that the radial entropy profile has a power-law behavior at large radii, $K\propto r^{\lambda_K}$, with $\lambda_K\approx1.1$. Later, full cosmological simulations \citep[][henceforth \citetalias{Voit05}]{Voit05} have shown that the radial entropy profile would be nearly constant near the center ($r\lesssim0.1\myR_{200}$), and approach a power-law profile at large radii ($r\gtrsim0.1\myR_{200}$), such that $K\simeq K_0+\bar{K}r^{\lambda_K}$. Here, $\myR_{x}$ is the radius enclosing a mean density $x$ times larger than the critical density of the Universe.
The power-law indices were fit as $\lambda_K=1.21\pm0.01$ for the SPH code, and $\lambda_K=1.24\pm0.03$ for the AMR code; henceforth we refer to this as $\lambda_K\approx1.2$, for brevity. After scaling the entropy of a GA of mass $M$ by the mass dependent factor $K_M\propto M^{0.26}$ \citepalias{Voit05}, $(\bar{K}/K_M)$ was found to vary among GAs by only $\lesssim3\%$, but different codes give constants $(K_0/K_M)$ that differ by a factor of $\sim2$.

Several studies of the observed entropy profiles in GAs were previously presented, as summarized in Table \ref{table:entropy fits}. Various methods were used. Some works analyzed the entropy $K$ as a function of radius $r$, while others studied the scaled entropy as a function of the scaled radius, $h(z)^{4/3}T^{-0.65}K=\Knorm(r/0.1\myR_{200})^{\lambda_K}$ \citep[\eg][]{Ponman03,Pratt05}.
Here, $z$ is the redshift, $h^2(z)\equiv\Omega_m(1+z)^3+\Omega_\Lambda$, and $\Omega_m$, $\Omega_\Lambda$ are the matter, vacuum energy fractions. Both pure radial power laws and power laws with a constant core, $K=K_0+K_{100}\left(r/100\kpc\right)^{\lambda_K}$, were considered. Some studies used the deprojected $T$, others fitted a parametric model for $T$, and some simply used the projected $T$.

\begin{deluxetable*}{cccccccccc}
\tablewidth{0pt}
\tablecaption{\label{table:entropy fits}
Studies of the entropy profile
}
\tablehead{
Reference & Sample Size & $M_{500}$ Range & Temperature & $K_0$ & $K_{100}$ & $\Knorm$ & $r [kpc]$ & $r/\myR_{200}$ & $\lambda_K$ \\
(1) & (2) & (3) & (4) & (5) & (6) & (7) & (8) & (9) & (10)
}
\startdata
\citetalias{Pratt05} & $5$ & $2\till8$  &  deprojected & -- & -- &$\sim470$&--&$0.01\till 0.8$& $0.94\pm0.14$ \\
\citetalias{Piffaretti05} & $13$ & $0.6\till8.7$ &  deprojected & -- &--&$504\pm140$&--&$0.01\till 0.8$& $0.95\pm0.02$ \\
\citetalias{Mahdavi05} & $8$ & $0.2\till 1.5$ & modeled & --&--&$\approx420$&--&$0.01\till 1$& $0.92^{+0.04}_{-0.05}$ \\
\citetalias{Pratt06} & $10$ & $0.2\till 1.5$ & modeled& -- &--&$\approx400	$ &--&$0.01\till 0.8$& $1.08\pm0.04$ \\
\citetalias{Donahue06} & $9$ & $1.5\till10$   & deprojected &  $7.3\pm4.6$ &$100\till210$&--& $1\till300$ &-- & $1.2\pm0.2$ \\
                       &     &                & deprojected & -- &$100\till210$&--&$1\till300$& --&$1.0\pm0.2$ \\
\citetalias{Sanderson09} & $9$ & $0.6\till12$& deprojected & --&--&$349\pm22$&--&$0.002\till 0.4$& $1.05^{+0.07}_{-0.15}$ \\
\citetalias{Johnson09} & $12$ & $0.3\till 1.3$ & projected&-- &--&$\approx330$&--&$0.007\till 1$& $0.71\pm0.02$ \\
\citetalias{Cavagnolo09} & $107$ & $0.2\till23$ & projected& $16.1\pm5.7$ &$100\till200$&--&$1\till 1000$&-- &$1.2\pm0.38$ \\
\citetalias{Pratt10} & $10$ & $1.7\till13$ & modeled & $25\pm25$&$120\pm55$&--&$20\till 1000$&-- &$0.8\till1.2$ \\
\citetalias{Panagoulia14} & $66$ & --  & deprojected & -- &$\sim240$&--&  $0.2\till100$ &-- & $\approx0.9$ \\
                          & $13$ &   & deprojected& -- &$\sim90$ &--&  $0.2\till20$ &-- & $\approx0.64$ \\
\hline
This work & $29$ & 0.05\till9 & deprojected &--&$187\pm6$&-- & $10\till 2000$ &-- & $0.96\pm0.01$ \\
\enddata

\tablecomments{
\textbf{Columns:} (1) Reference abbreviations: \citetalias{Pratt05}---\citet{Pratt05}, \citetalias{Piffaretti05}---\citet{Piffaretti05}, \citetalias{Mahdavi05}---\citet{Mahdavi05}, \citetalias{Pratt06}---\citet{Pratt06}, \citetalias{Donahue06}---\citet{Donahue06}, \citetalias{Sanderson09}---\citet{Sanderson09}, \citetalias{Johnson09}---\citet{Johnson09}, \citetalias{Cavagnolo09}---\citet{Cavagnolo09}, \citetalias{Pratt10}---\citet{Pratt10}, \citetalias{Panagoulia14}---\citet{Panagoulia14}; (2) number of GAs in the sample; (3) mass range of the sample, within $\myR_{500}$ in $10^{14}M_\odot$ units; (4) method of temperature extraction (see the text); (5) the core value $K_0$ $[\mbox{keV}\cm^{-2}]$; (6) the constant $K_{100}$ $[\mbox{keV}\cm^{-2}]$, for non-scaled fits; (7) the constant $\Knorm$ $[\mbox{keV}\cm^{-2}]$, for scaled fits; (8) radial range in kpc for non-scaled fits; (9) radial range in $\myR_{200}$ for scaled fits; (10) entropy power-law index $\lambda_K=d\log K/d\log r$.}
\end{deluxetable*}

In general, results that are not fully deprojected should be treated with caution, as the projection effects \citep[\eg][]{Panagoulia14} and the exact model fitted \citep[\eg][]{Pratt06} can have a significant effect on the results. This can explain some but not all of the apparent inconsistencies in the table.
The deprojected samples, when averaged, give $\lambda_K\sim(0.9\till1.1)$. \cite{Panagoulia14} found no evidence for an `entropy floor' in the centers of GAs,
and suggested that the floor found by others \citep[\eg][]{Cavagnolo09} may be due to a combination of resolution and projection effects.
No redshift dependence was found in the $z\in[0.1\till1]$ range within $r<0.7\myR_{500}$ \citep{Pratt10}.

In the past decade, high resolution X-ray observations have revealed the presence of spiral patterns in a significant fraction of the CC clusters
for which high quality data are available \citep[\eg][]{Clarke04, Tanaka06, Sanders09, Johnson10, Lagana10, Randall10, Blanton11, Roediger11,Nulsen13,Venturi13,Giacintucci14,Ghizzardi14,Sanders14}. These observations suggest a spiral morphology composed of spatially alternating, low entropy and high entropy plasma phases, which are approximately at a local pressure equilibrium.
Recent, deep observations found hints of spiral patterns in clusters on a much larger, Mpc scale \citep{Simionescu12, Rossetti13, Walker14}.
Lately, deep observation of CC groups revealed similar features \citep{Gastaldello09,Randall09,Machacek11,Canning13,Gastaldello13}.

The boundary between the low entropy phase from below (\ie closer to the center of the GA), and the high entropy phase from above, may form a discontinuity, observed as an X-ray edge known as a cold front (CF) where projection effects are favorable.
In CCs, CFs are the telltale signs of an underlying spiral structure.
Indeed, such CFs are often quasi-spiral, piecewise spiral, or nearly concentric, suggesting an underlying three-dimensional (3D) spiral pattern seen in various projections. Multiple, quasi-concentric CFs are often found on alternating sides of the center of the GA, with an increasing distance and size, consistent with a spiral discontinuity manifold observed edge on. Examples include the clusters A2142, RXJ1720.1+2638, A2204, and A496 \cite[][and references therein]{Markevitch07}, and the group NGC5098 \citep{Randall09}.

As a local phenomenon, CFs are easier to detect and quantify than spiral patterns.
Hence, CFs provide the main evidence for the prevalence of spiral structures in GAs.
CFs were found in more than half of the otherwise relaxed CCs \citep{Markevitch03}, and nearly in all of the well-observed, low redshift cores. For example, at least one CF was found in each of the 10 CCs in the sample of \citet{Ghizzardi10}.

CFs are particularly useful for studying spiral patterns because they constrain the underlying dynamics.
Indeed, such CFs typically show deviations from hydrostatic equilibrium \citep[\eg][]{Mazzotta01} that reflect fast flows and shear below the CF \citep{Keshet10}.
The pressure jumps resulting from shear magnetization, predicted in CC CFs both analytically \citep{Keshet10} and numerically \citep{ZuHone11}, were recently measured at the $\sim10\%$ level \citep{Reiss14}.
The low entropy component below the CF is colder and denser (both by a similar factor $q$ of up to a few) and higher in metallicity than the plasma above it \citep[\eg][]{Markevitch07, Ghizzardi14}.

The spiral pattern of CFs and the fast flow below them, when combined, indicate the presence of a spiral flow.
Note that by 'spiral flow' we refer to a plasma distribution giving rise to spiral features; Lagrangian fluid elements do not necessarily follow spiral trajectories \citep{Ascasibar06}.
The ubiquity of spiral patterns and CFs suggests that such spiral flows constitute an inherent part of the GA itself. Indeed, a simple spiral flow model reproduces the main thermal features of CCs, such as the density cusp, while avoiding the local thermal instability by advecting heat across and beyond the core.
The model is also consistent with the nearly logarithmic spiral pattern, the properties of the projected CFs, and the fast flow along them \citep[][henceforth \citetalias{Keshet12}]{Keshet12}. However, the recent evidence for extended, Mpc scale spiral patterns suggests that spiral flows are, in fact, a GA-wide phenomenon.

We study the two universal properties of GAs discussed above in conjunction, namely a universal power-law entropy, or entropy-like, profile (henceforth: universal profile) and an underlying, extended spiral flow, reaching beyond the core. After evaluating the universal profile and arguing that it is likely to arise from a dynamical mechanism, we point out that a GA-wide spiral flow can naturally explain it, thus linking the two phenomena.

The paper is organized as follows. In Section \ref{sec:Scaling} we analyze the entropy profiles of deprojected GAs found in the literature. We also study a different, but nearly indistinguishable measure of the thermal profile, namely the ratio between the cooling and free-fall times. In Section \ref{sec:dynamic} we argue for a dynamical origin of the universal profile. We show how it naturally arises in a spiral flow, using two different arguments, in Section \ref{sec:spiral origin}. In Section \ref{sec:Spiral} we demonstrate this using a third, more rigorous argument. Here we generalize the two-phase CC spiral flow to span the entire GA, surprisingly reproducing the universal profile. Our results are summarized and discussed in Section \ref{sec:Discussion}.

We adopt the $\Lambda$CDM cosmological model with a Hubble constant $H_0=70\km\se^{-1}\Mpc^{-1}$. Assuming a $76\%$ hydrogen mass fraction gives a mean particle mass $\mu=0.59m_p$ and a particle number density $n=\zeta n_e$, where $m_p$ is the proton mass and $\myzeta\simeq 2.1$.
Confidence intervals are $68\%$ for one parameter.
All fit uncertainties were estimated from the dispersion of the data points around the best fit, because propagating the measurement uncertainties of each data point makes a small (by a factor of a few) contribution to the total uncertainty.

\section{A universal thermal profile}
\label{sec:Scaling}

\subsection{Specific Entropy Profile}
\label{subsec:Scaling_SpecificEntropyProfile}
We examine the azimuthally averaged, deprojected radial thermal profiles of GAs from the literature.
The GAs analyzed are summarized in Table \ref{table:my data table}.
The sample is composed of GAs with a relatively low redshift, $z<0.25$.
They span a wide range of masses, $M_{500}=[0.05-9]\times10^{14}M_\sun$, where $M_{500}$ is the mass enclosed inside $\myR_{500}$.

We compute $K$ for each radial data bin in each GA.
Following \cite{Panagoulia14}, we remove the innermost data point in each GA in order to reduce resolution effects.
The number density $n_e$ is derived from the deprojected X-ray surface brightness, $\SB$.

The temperature $T$ is derived using two different methods: (i) by deprojecting the same X-ray data used for $n_e$; or (ii) by combining $n_e$ derived from ROSAT $\SB$ maps, with the electron thermal pressure, $P\equiv\kB T \zeta n_e$, derived from the Sunyaev--Zel'dovich (SZ) effect measured by \emph{Planck} \citep{Eckert12, Eckert13, Planck5}.
These two methods agree well where they overlap (A1795, A2029, A3112, A2052, and A2204), showing that the systematic errors in the temperatures derived are small.

\begin{deluxetable}{lccc}
\tablecaption{\label{table:my data table}
Sample details.
}
\tablecolumns{4}
\tablewidth{0pt}
\tablehead{
\colhead{GA Name} & \colhead{Deprojection Reference} & \colhead{\emph{Planck} SZ} & \colhead{$M_{500}$} \\
\colhead{(1)} & \colhead{(2)} & \colhead{(3)} & \colhead{(4)}
}
\startdata
2A0335 & \cite{Kaastra04} & -- & 1.62\\
A0085 & -- & \cmark & 5.06 \\
A0262  & \cite{Kaastra04} & -- & 0.68\\
       & \cite{Sanders10} &  &\\
A0399 & \cite{Kaastra04} & -- & 7.44 \\
A0478 & -- & \cmark & 8.42 \\
A0496  & \cite{Kaastra04} & -- & 2.04\\
A1795  & \ibid & \cmark & 7.21\\
A1835  & \ibid & -- & 6.33\\
A1837  & \ibid & -- & 1.46\\
A2029  & \cite{Lewis03} & \cmark & 8.67\\
A2052  & \cite{Kaastra04} & \cmark & 1.42\\
A2204  & \cite{Sanders09}\tablenotemark{a} & \cmark & 6.44\\
A3112  & \cite{Bulbul12} & \cmark & 3.83\\
A3581  & \cite{Sanders10} & -- & 0.70\\
A4059  & \cite{Kaastra04} & -- & 2.93\\
AWM4   & \cite{Osullivan10} & -- & 1.04\\
HCG62 & \cite{Gitti10} & -- & 0.29\\
      & \cite{Sanders10} &  & \\
M49 & \cite{Kraft11} & -- & 0.94\tablenotemark{b} \\
MKW9  & \cite{Kaastra04} & -- & 0.88 \\
NGC0533 & \ibid & -- & 0.32 \\
NGC0741 & \cite{Jetha08} & -- & 0.70 \\
NGC4261 & \cite{Osullivan11} & -- & 0.77 \\
NGC5813 & \cite{Randall11} & -- & 0.18 \\
NGC4325 & \cite{Russell07} & -- & 0.23 \\
NGC4636 & \cite{Werner09} & -- & 0.11 \\
NGC6482 & \cite{Khosroshahi04} & -- & 0.05\\
Perseus& \cite{Churazov03} & -- & 4.97 \\
       & \cite{Urban14} &  & \\
RXJ1159 & \cite{Humphrey12} & -- & 0.59  \\
Virgo  & \cite{Kaastra04} & -- & 0.94  \\
\enddata
\tablecomments{\textbf{Columns:} (1) GA name; (2) deprojected temperature data reference; (3) existence of SZ data \citep{Eckert12, Eckert13, Planck5}; (4) cluster mass $[10^{14}M_\odot]$ \citep{Reiprich02,Osmond04,Piffaretti05,Pratt05,Finoguenov06,Gastaldello09b,Eckmiller11,Urban11,Humphrey12}.}
\tablenotetext{a}{Showing the four sectors separately.}
\tablenotetext{b}{Mass of the host cluster, Virgo.}
\end{deluxetable}

\begin{figure}
\centerline{\epsfxsize=8.4cm \epsfbox{\myfig{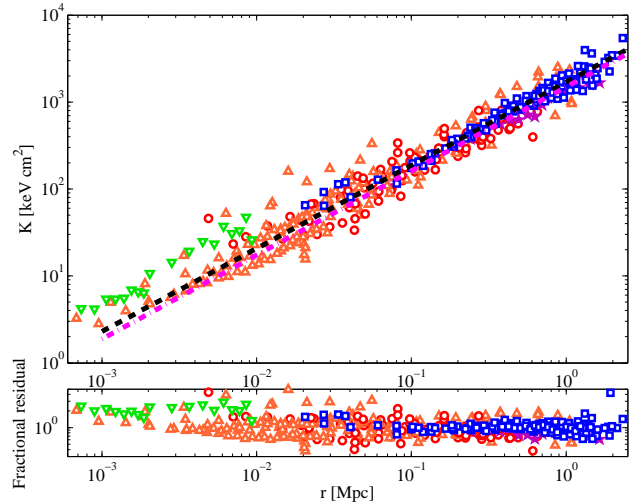}}}
\caption{\label{fig:entropy scaling} Top: specific entropy $K$ as a function of radius, for all radial bins of all objects in the sample (Table \ref{table:my data table}). The temperature used to compute $K$ was derived either from the deprojection of X-ray spectra (red circles: \emph{XMM-Newton}; orange triangles: \emph{Chandra}; purple pentagram: \emph{Suzaku}) or from the SZ pressure (blue squares).
Also shown are data for M49 (green down triangles), the $r>10\kpc$ power-law fit for the entire data (dashed black; Equation \eqref{eq:K fit}), and the weighted mean over individual GA fits (dotted-dashed magenta; Equation \eqref{eq:K fit_per_GA}). Bottom: the fractional residuals from the power-law fit to all the data; Equation~(\ref{eq:K fit}).}
\end{figure}

In Figure \ref{fig:entropy scaling} the entropy $K$ is plotted as a function of radius for all the GAs of our sample. For clarity, individual measurement uncertainties are not shown. The results are well-fit $(1-R^2\simeq 0.072)$ in the $r\simeq[0.01\till2]\Mpc$ range by a pure power law,
\begin{equation}  \label{eq:linear_entropy}
K=K_{100}\left(r/100\kpc\right)^{\lambda_K} \, ,
\end{equation}
with
\begin{equation}
\label{eq:K fit}
\lambda_K=0.96\pm0.01; \quad \mbox{and} \quad K_{100}=187\pm6\keV\cm^2 \, ,
\end{equation}
and with a fractional dispersion $\sigma(\log_{10}K_{100})\simeq 0.4$.
Deviations from this power-law are seen for $r\lesssim10\kpc$, where $K(r)$ slightly flattens.

Using a smaller radial extent, $[0.3\till1.5]R_{500}$ \citepalias[similar to][]{Voit05}, gives $\lambda_K=1.00\pm0.05$.
Fitting the scaled entropy to the scaled radius gives $\lambda_K=0.96\pm0.01$ for $r>10\kpc$, but the fit is not as good ($1-R^2\simeq 0.12$) as it is for the non-scaled fit.

\begin{figure}
\centerline{\epsfxsize=8.4cm \epsfbox{\myfig{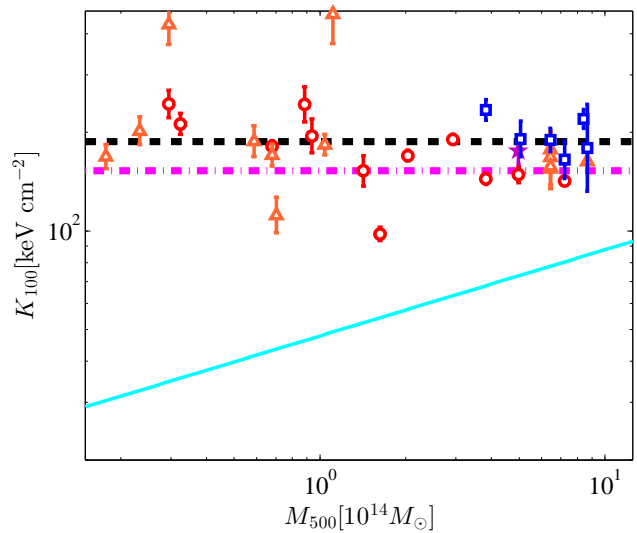}}}
\caption{\label{fig:K100 by cluster} Specific entropy at $100\kpc$, $K_{100}$, plotted as a function of $M_{500}$. The symbol and line notations are the same as in Figure \ref{fig:entropy scaling}. Also shown (solid cyan) is the entropy of the non-radiative self-similar model for $z=0$ \citepalias{Voit05}}.
\end{figure}

Each GA can be separately fit with a power-law profile.
The (uncertainty weighted) mean over all GAs then gives
\begin{equation}
\label{eq:K fit_per_GA}
\lambda_K^{sep}=0.97\pm0.01; \quad \mbox{and} \quad K_{100}^{sep}=153\pm2\keV\cm^2\, ,
\end{equation}
but there is a considerable scatter, $\sigma(\lambda_K)=0.20$ and $\sigma(K_{100})=72\keV\cm^2$, among different GAs.
The scatter in $K_{100}$ is largely due to two outlier GAs; see Figure~\ref{fig:K100 by cluster}.

\begin{figure}
\centerline{\epsfxsize=8.4cm \epsfbox{\myfig{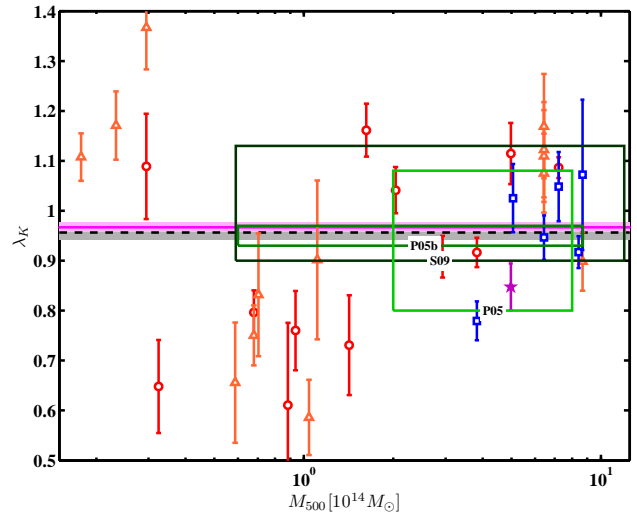}}}
\caption{\label{fig:entropy by cluster} Specific entropy power-law index $\lambda_K$, as a function of $M_{500}$, in individual GAs. The symbol and line notations are the same as in Figure \ref{fig:entropy scaling}. Also shown (as labeled rectangles) are the fits of \citet{Pratt05}, \citet{Piffaretti05}, and \citet{Sanderson09}.
}
\end{figure}

Figures \ref{fig:entropy by cluster} and \ref{fig:K100 by cluster} show $\lambda_K$ and $K_{100}$, respectively, as a function of the GA mass, for individual GAs with sufficient data (more than 4 data points beyond $10\kpc$).
Neither $\lambda_K$ nor $K_{100}$ show a correlation with $M_{500}$.
Note that a $K_{100}\propto M^{0.26}$ correlation would be expected in the non-radiative, accretion-only models \citepalias{Voit05} discussed above.

As these results show, the uncertainty in the power-law index $\lambda_K$, if the latter is assumed to be a universal constant, is very small and dominated by the inherent dispersion.
The power-law fits (Equations \eqref{eq:K fit} and \eqref{eq:K fit_per_GA}) are shown in Figures \ref{fig:entropy scaling}--\ref{fig:K100 by cluster} as dashed and dotted-dashed lines.

It is interesting to examine if the above universal thermal profile also applies to giant elliptical galaxies that harbor an AGN but do not lie at the center of a GA.
Deprojected density and temperature profiles from \emph{Chandra} are available \citep{Kraft11} for the giant elliptical M49, which resides in the outskirts of the Virgo cluster.
The $K$ and $\tau$ profiles around the galaxy, shown in Figures.~\ref{fig:entropy scaling} and \ref{fig:universal scaling}, do indeed resemble the universal scaling.
Note that the M49 data were not included in the fit formulae above, as they do not meet the $r>10\kpc$ selection criterion.

Recent \emph{Suzaku} observations revealed a flattening in the observed entropy profile at large radii \cite[$r\gtrsim 1\Mpc$; \eg][]{Walker12}. This departure from a power-law behavior has been commonly attributed to gas clumping, which leads standard X-ray modeling to overestimate the gas density and thus underestimate the entropy. No significant flattening is seen in the data shown in Figure~\ref{fig:entropy scaling} (see also Appendix \ref{sec:flat}), possibly because the effect is small in the relevant radial range \citep[see, \eg][]{Eckert13Clump}, or due to some bias in the \emph{Suzaku} observations \citep[see, \eg][]{MorandiEtAl15}. Restricting the analysis to small radii, in the range $10\kpc<r<1\Mpc$, gives $\lambda_K=0.94\pm0.01$; slightly \emph{flatter} than Equation~\eqref{eq:K fit}.

\subsection{Cooling Time to Free-fall Time Ratio}

There are other ways to quantify the universal thermal profile of GAs.
Following \cite{McCourt12} and \cite{Sharma12}, we examine the ratio $\tau\equiv \tcool/\tff$ between the cooling time $\tcool$ and the free-fall time $\tff$ of the IGM.

We define $\tcool$ similarly to \cite{McCourt12}, as the ratio between the thermal energy density and the cooling rate of the X-ray emitting gas,
\begin{equation}
\tcool \equiv \frac{U_{th}}{dE/dt} = \frac{(3/2)\myzeta n_e\kB T}{n_e^2\Lambda(T)} \, .
\end{equation}
Here $\Lambda(T)$ is the cooling function, which we adopt from \cite{Tozzi01}, with a nominal 0.3 solar metallicity.

For simplicity, and to minimize the uncertainties, we define the free-fall time differently than \cite{McCourt12}, using the constant density approximation,
\begin{equation}
\label{eq:tff def}
\tff \equiv (G\rhotot)^{-1/2} = (\fb/G\rho)^{1/2}\, .
\end{equation}
Here $G$ is Newton's constant, $\rho=\mu\myzeta n_e$ is the mass density of the X-ray emitting gas, and $\fb\equiv\rho/\rhotot$ is the gas mass fraction of the IGM, of order $[2\till15]\%$ for $r/\myR_{500}>0.01$ \citep{Eckert13II}.
Alternative definitions of $\tff$ differ from the above by a constant factor of order unity, which we omit; this bears no consequences for the following discussion.

With the above definitions, the ratio between the two timescales becomes
\begin{equation}
\tau\propto \left[ \frac{T^{1/4}}{\fb^{1/2}\Lambda(T)} \right] K^{3/4} \, .
\end{equation}
The prefactor in the square parenthesis does not vary significantly with $r$.
Hence, one may expect $\tau$ to scale roughly as $K^{3/4}\propto r^{3/4}$.

\begin{figure}[h]
\centerline{\epsfxsize=8.4cm \epsfbox{\myfig{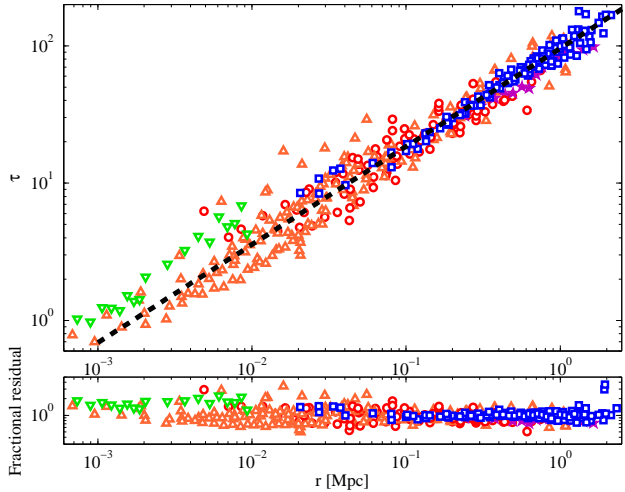}}}
\caption{\label{fig:universal scaling}
Top: the ratio $\tau$ between the cooling time and the free-fall time, plotted as function of radius. Bottom: the fractional residuals relative to the fit Equation~\eqref{eq:simple fit}.}
\end{figure}

The gas fraction $\fb$ is not well constrained,  but is not thought to vary over more than an order of magnitude across a GA. As a first approximation, we thus assign it with a constant, $\fb=0.15$ value, and refine this assumption below.
Indeed, $\tau$ is then well-fit by a pure power law in the range $r\simeq[0.01\till2]\Mpc$,
\begin{equation}
\label{eq:simple fit}
\tau= \left(3.6\pm0.1 \right) r_{10}^{0.72\pm0.01}\, ,
\end{equation}
where $r\equiv (r/10\kpc)$, as shown in Figure~\ref{fig:universal scaling}.
The dispersion of the fractional residuals to the fit is $\sigma=0.3$.
This power-law behavior slightly flattens for $r\lesssim10\kpc$, which is the radius where $\tau\simeq 1$.

Using the high temperature approximation for the cooling function, $\Lambda(T)\simeq\Lambda_0 T^{1/2}$, where $\Lambda_0\simeq2.1\times10^{-27}\erg \se^{-1} \cm^{3} \mbox{ K}^{-1/2} $
\citep[\eg][]{Rybicki}, Equation \eqref{eq:simple fit} can now be crudely written as
\begin{equation}
\label{eq:conduction scale}
C\frac{(\kB T)^{2/3}}{n_e^{2/3}r} \simeq
\frac{\tau^{4/3}}{r}
\simeq \frac{r_{10}^{0.04}}{L_\tau}
\approx L_\tau^{-1} = (2\kpc)^{-1} \, ,
\end{equation}
where $C\equiv [9G \kB\mu\myzeta^3/(4\fb\Lambda_0^2)]^{2/3}$ is a constant under the present (constant $\fb$ and $\Lambda/T^{1/2}$) assumptions, and we defined $L_\tau\simeq 2\kpc$ as the constant thus measured from $(\tau^{4/3}/r)$.
Equation \eqref{eq:conduction scale} resembles the entropy fit (Equations~\eqref{eq:linear_entropy} and \eqref{eq:K fit}), which may analogously be written in the form
\begin{equation}
\label{eq:entropy scale}
\frac{\kB T}{n_e^{2/3}r} = \frac{K}{r} \simeq 1.0\times10^{-30}r_{10}^{-0.04}\erg\cm
\simeq 10^{-30}\erg\cm \, .
\end{equation}

Equations (\ref{eq:conduction scale}) and (\ref{eq:entropy scale}) show that the observed GA thermal profiles are consistent with a universal constant value of either $(\tau^{4/3}/r)\propto T^{2/3}/(n_e^{2/3}r)$, or $(K/r)\propto T/(n_e^{2/3}r)$.
These two quantities can be distinguished only by their slightly dissimilar temperature dependence, differing by a factor $T^{1/3}$.
With the present data, it is difficult to determine which of the two quantities better describes GAs because the temperature range in each GA is quite narrow (usually less than an order of magnitude), the temperature differences between the GAs in our sample are not large, and the temperature uncertainties are considerable.
Indeed, the fits in Equations \eqref{eq:K fit} and \eqref{eq:simple fit} are presently of similar qualities ($1-R^2=0.072$ and $0.070$, respectively).

Equations~(\ref{eq:simple fit}) and (\ref{eq:conduction scale}) were derived by approximating $\fb$ as a constant.
This can be relaxed using estimates of $\fb$ from the literature.
Using the average radial dependence of $\fb$ \citep[median values]{Eckert13II} gives a poorer fit ($1-R^2=0.094$) to the data.
Using the estimated mass dependence of $\fb$ \citep{Gonzalez13} renders the fit even worse ($1-R^2=0.129$).
In both cases, the quantity $(\tau^{4/3}/r)$ is found to be nearly constant, with a radial dependence weaker even than the mild $r_{10}^{0.04}$ of Equation~\eqref{eq:conduction scale}.

Note that the dimensional constants on the right sides of Equations~\eqref{eq:conduction scale} and \eqref{eq:entropy scale} are not naturally obtained from the constants that appear in the fluid equations (namely, the continuity, Euler, and energy equations; see section \ref{sec:Spiral}), without introducing a large number or some new dimensional constant, even if conduction and cooling are taken into account.
For example, one may construct the length scale $L_0\equiv\Lambda_0G^{-1}k_B^{-1/2}\mu^{-3/2}\simeq 1$ Mpc from the constants in these equations, but this is larger by a factor of $\sim 500$ than $L_\tau$.
It would be natural if the cutoff on the thermal correlation itself (a few kpc; see Equation~\eqref{eq:conduction scale}) would somehow provide the necessary scale.

The simple power-law function $\tau(r)$ that we find, as shown in Equation.~\eqref{eq:simple fit} and Figure \ref{fig:universal scaling}, differs from the results of \cite{McCourt12} in two ways.
First, we find that $\tau$ monotonically decreases toward small radii rather than showing a minimum, probably because we use only deprojected data, adopt a more robust definition of $\tff$, and avoid a parametric model fit.
This is consistent with the absence of an `entropy floor' in our data, in accordance with \cite{Panagoulia14}. Second, we find a universal behavior for all the GAs examined, regardless of the presence of H$\alpha$ filaments.

\section{Dynamic origin of the universal profile}
\label{sec:dynamic}

The universality of the thermal profile discussed in Section \ref{sec:Scaling}, its extensive radial span, and its applicability over a wide mass range, place strong constraints on its origin.
As we show, the static models proposed so far seem unable to reproduce the profile without fine tuning.

The entropy seeded by primordial gas crossing the virial shock \citep{Tozzi01} would not naturally maintain the same power-law profile down to and beyond the core radius, where cooling plays a significant role on timescales shorter than the age of the GA.
Rather, one would expect some break in the profile, unless some additional mechanism compensates for the radiative entropy loss.
Whether or not such a mechanism exists, the primordial entropy should scale with the GA mass as $K_{100}\propto M_{500}^{0.26}$ \citepalias{Voit05}, but we find no such correlation (see Figure \ref{fig:K100 by cluster}). Rather, the power-law index of the $K_{100}-M_{500}$ correlation has a slightly negative best-fit, is consistent with zero ($<2\sigma$), and is inconsistent with the predicted self similar value by $>8\sigma$ (this applies to both SPH- and AMR-based fits of \citetalias{Voit05}; henceforth). This result does not change much if the representative entropy is chosen well outside of the core (\eg $\gtrsim 5\sigma$ at $500\kpc$, and $\gtrsim4\sigma$ at $1000\kpc$). Moreover, as noticed previously \citep[\eg][]{Pratt10}, the measured value of the entropy lies well above the prediction of the self-similar model, over the entire mass range (see Figure \ref{fig:K100 by cluster}). This deviation ranges between $1\sigma$ and $20\sigma$ for individual clusters, and is $>6.5\sigma$ for their mean. Finally, the entropy profile we find is significantly flatter (by $>8\sigma$) than the self-similar $\lambda_K\approx1.2$, which can be ruled out in our sample. Even by using a much smaller radial extent for the fit ($[0.3\till1.5]\myR_{500}$, similar to \citetalias{Voit05}), the self-similar prediction is still rejected by $>4\sigma$.
These arguments reject self-similar primordial accretion alone as the origin of the universal profile, and indicate a pervasive baryonic influence reaching out to the virial radius.

Another previous suggestion is that the entropy profile is governed by radial heat conduction \citep{Zakamska03,Dolag04}.
This mechanism is unlikely to operate in exactly the same fashion on both sides of the temperature peak, where the radial direction of the heat flow is reversed, so again the thermal profile would naturally be broken.
Moreover, \cite{Sanderson09} showed that in order for conduction to thus explain the GA thermal profile, the heat conductivity should have a fine-tuned, unnatural functional form.

More generally, the IGM is a dynamical environment, undergoing merger activity \citep[\eg][]{Vikhlinin14}, susceptible to the significant energy output of the AGN \citep[\eg][]{Birzan04}, and harboring ongoing shocks \citep[\eg][]{Markevitch07}, fast bulk flows \citep{Keshet10}, and turbulence \citep[\eg][]{Zhuravleva14}.
Static models are unlikely to robustly explain the universal profile in such surroundings.

In contrast, a dynamical mechanism, involving gas motions, could have the inertia to resist the turmoil of the IGM and stabilize the universal profile.
Such a mechanism would have to operate throughout the GA, out to the virial radius, in order to modify the (mass-correlated) entropy injected by the virial shock, and to maintain the universal profile.
It would also need to be insensitive to the sign of the radial temperature gradient in order to produce the same profile both inside and outside the core.

If the universal quantity turns out to be $(\tau^{4/3}/r)$, rather than $(K/r)$, this would provide a direct indication that the process is dynamic, and is, at least on small scales, driven by gravity.
The length scale $L_\tau\simeq 2\kpc$ emerging from this quantity when applied to GAs (Equation~\eqref{eq:conduction scale}) would suggest a non-local effect, such as an extended bulk flow emanating from this radius.

\section{Spiral flow origin of the universal profile}
\label{sec:spiral origin}

As demonstrated by \citetalias{Keshet12}, spiral flows can explain the thermal structure of cores. It is natural to ask, in light of the spiral features found to extend out nearly to the virial radius, if a spiral flow reaching beyond the core can give rise to the universal profile. This is crudely examined here, and examined in more detail in the specific context of a two-phase spiral flow model in Section \ref{sec:Spiral}.

Consider a two-dimensional spiral flow. The more complicated, 3D case can be approximated under some assumptions as a superposition of spiral flow planes (see \citetalias{Keshet12} and Section~\ref{sec:Spiral} below). The spiral flow necessarily involves azimuthal density and temperature gradients, as evidenced by the jumps in $n_e$ and $T$ across the spiral CFs. As we show, the universal profile can be recovered by making simple, although ad hoc, assumptions regarding these gradients.

\begin{figure}
\centerline{\epsfxsize=8.4cm \epsfbox{\myfig{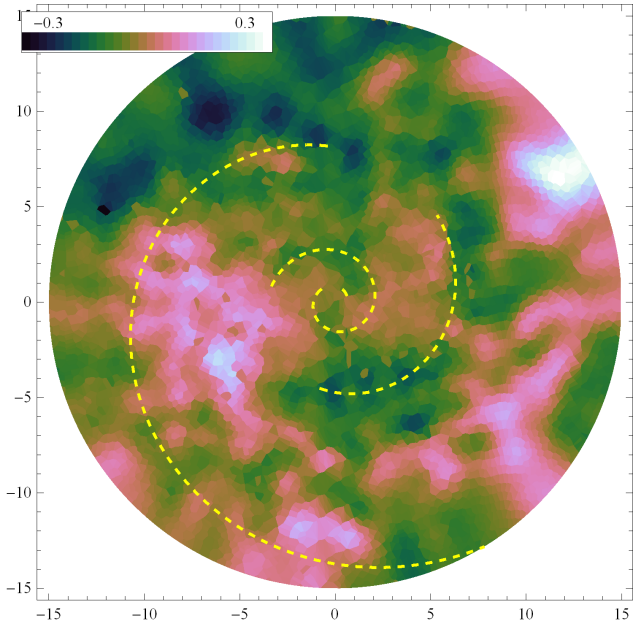}}}
\caption{
\label{fig:PerseusPressure}
Fractional deviation of the pseudo-pressure $\tilde{P}=\SB^{1/2} T$ from its azimuthal average in a 30' by 30' region around the center of Perseus (adopted from the \cite{Churazov03} analysis of \emph{XMM-Newton} data, courtesy E. Churazov; color scale from \cite{Green11}).
The best-fit spiral of \citetalias{Keshet12} is shown (dashed yellow) based on a gradient filter applied to $\SB$.
}
\end{figure}

First note that in order to sustain a spiral flow, mild azimuthal pressure gradients are also necessary (\citetalias{Keshet12}). These can be seen, for example, in Figure \ref{fig:PerseusPressure}, which shows the fractional deviation of the pseudo-pressure $\tilde{P}=\SB^{1/2}T$ from its azimuthal average, in Perseus \citep{Churazov03}. In spite of the projection and noise, a spiral structure is clearly seen. It is highlighted by the overplotted spiral CF pattern, derived \citepalias{Keshet12} based on thresholding the $\SB$ gradients.
Indications for azimuthal pressure gradients are also seen in deprojected pressure profiles of CC GAs \citep{Sanders09,Urban14}.

\subsection{Scaling of the azimuthal gradients}

Next, consider the normalized azimuthal derivative $\partial \log A/\partial\phi$ of some thermal quantity $A\in\{T,P,K\}$, and in particular its radial dependence.
In general, such derivatives tend to decrease with increasing radius, as the spiral sections of the flow become broader.
The precise scaling would depend on the spiral pattern and on the nature of the flow.
For simplicity, we assume that there exists some part of the flow in which
\begin{equation} \label{eq:spiral_scaling}
\partial_\phi \log A\propto 1/r \, .
\end{equation}

To see how such a scaling could locally emerge, parameterize the geometric pattern of a 2D spiral by $\lambda_\gamma\equiv d\log|\gamma|/d\log(r)$, where $\gamma\equiv \tan(\alpha)$, and $\alpha$ is the angle between the streamline and the azimuth (\citetalias{Keshet12}, correcting a typo in the definition of $\gamma$ therein). Thus, $\lambda_\gamma=0$ corresponds to a logarithmic spiral, while $\lambda_\gamma=-1$ gives an Archimedean spiral. Observations, simulations, and the core spiral model suggest a nearly logarithmic spiral, with a slight Archimedean tendency, $-1<\lambda_\gamma<0$ \citepalias{Keshet12}.
If the spiral structure is separable, $A=A_r(r)A_\phi(\phi-\phi_0(r))$, then $\partial \log A/\partial\phi\sim 1/\log(r)$ in a simple logarithmic spiral, and $\partial \log A/\partial\phi\sim 1/r$ in a simple Archimedean spiral.

However, neither of these spiral structures, nor any other uniform $\phi$ derivative, can apply globally (\ie for $0<\phi<2\pi$) because the CFs observed indicate that a convective layer must be embedded in the spiral structure.
To see this, note that a spiral CF implies both radial and azimuthal entropy gradients. A generalized Schwarzschild criterion for convective stability then shows that a convective region must be present at any radius, as shown in Appendix \ref{sec:Schwarzschild}.
An alternative origin for Equation.~(\ref{eq:spiral_scaling}) would be a quasi 3D flow near the convective layer, with an effectively 2D gradient.

As the spiral structure cannot be monotonic, a simple analytic description of the entire flow is not feasible.
Some regions of the spiral should, however, show a radially declining $\partial \log A/\partial \phi$ derivative.
For simplicity, in the remainder of this section we assume that there exists some not necessarily large region of the flow in which $T$ or $P$ follow Equation~(\ref{eq:spiral_scaling}), such that $\partial_{\phi}\log T\sim L/(2\pi r)$ or $\partial_{\phi}\log P\sim L/(2\pi r)$.
Here $L$ is some constant length scale, discussed below.

\subsection{Cooling balanced locally either by azimuthal heat conduction or radial heat advection}

Under the above assumption, consider a radiatively cooling part of the flow.
The spiral structure brings hot and cold plasma phases near each other, so azimuthal heat conduction can balance the cooling.
This would imply that\footnote{We consider the mixed $r,\phi$ derivative terms in the energy (Equation~\eqref{eq:energy}), as
the lowest order (in $r/L$) azimuthal terms.}
$(\kappa/r)\partial_{\phi}T\sim r\Lambda_0 n_e^2T^{1/2}$, so
\begin{equation} \label{eq:Spiral_K}
\frac{K}{r}=\frac{\kB T}{n_e^{2/3}r}\simeq \kB\left(\frac{2\pi \Lambda_0}{\kappa_0L}\right)^{1/3}= \text{const.,}
\end{equation}
where $\kappa\simeq \kappa_0T^{5/2}$ is the Spitzer conductivity, with $\kappa_0\simeq10^{-6}\mbox{ erg s}^{-1}\mbox{cm}^{-1}\mbox{K}^{-7/2}$.
This is precisely the form of the universal profile (Equation \eqref{eq:entropy scale}), and would give the correct constant provided that $L\simeq 10$ kpc.

Alternatively, consider the advection of heat in a spiral flow, either inward from the heat bath outside the core or outward from some energetic outflow originating deep within it.
To find the radial velocity $\myv$, we assume the this heat flow is aligned with the spiral pattern, $\myv=\gamma\myw$, where the azimuthal velocity $\myw$ may be found from momentum conservation in the azimuthal direction, $r^{-1}\myw^2\sim(\gamma r\rho)^{-1}\partial_{\phi}P$.
These relations are rigorously derived in Section \ref{sec:Spiral}, in a frame rotating with the spiral pattern.

A balance between the radially advected heat and the radiative cooling would thus imply that $\Lambda_0 n_e^2T^{1/2}\sim(3/2)\zeta n_e\kB Tv/r$, so
\begin{equation} \label{eq:Spiral_tRatio}
\frac{1}{L_\tau} \simeq \frac{\tau^{4/3}}{r}
\simeq C\frac{(\kB T)^{2/3}}{n_e^{2/3}r}
\simeq \left(\frac{9\pi \zeta^4}{2\gamma\fb^2}\right)^{1/3}\frac{1}{L_0^{2/3}L^{1/3}} = \text{const.,}
\end{equation}

where the first line restates Equation~(\ref{eq:conduction scale}), and we used the estimate $\gamma\simeq 0.2$ \citepalias{Keshet12}.
This is precisely the form of the universal profile Equation \eqref{eq:conduction scale} and would give the correct constant, provided that $L\simeq 1$ kpc.

The estimates (Equations~\eqref{eq:Spiral_K} and \eqref{eq:Spiral_tRatio}) are accurate only to an order of magnitude, due to the unknown geometric factors. They depend on our ad hod assumption regarding the azimuthal gradient. However, it would suffice if this assumption holds only for some part of the flow that follows the universal scaling, such as the region on one side of the CF or of the convective layer. Any such layer with $\partial_\phi \log P\propto 1/r$ flowing along the spiral pattern or with $\partial_\phi \log T\propto 1/r$ would do.

A priori, these two independent estimates of $L$ did not have to agree.
In particular, the former does not depend on $G$, and the latter has nothing to do with $\kappa$.
Their agreement, and the plausible value of $L$ as the base radius of the spiral and as the radius where the universal profile flattens, suggests that such conduction-supported flows may have merit. Furthermore, this supports a dynamical origin of the universal profile in Equations~\eqref{eq:conduction scale} and \eqref{eq:entropy scale} and provides hints for a more sophisticated spiral flow modeling.

Next, we recover the universal profile from a more detailed, specific model of a spiral flow.

\section{Two-Phase Spiral Model}
\label{sec:Spiral}

\begin{figure}
	\centerline{\epsfxsize=7.5cm \epsfbox{\myfig{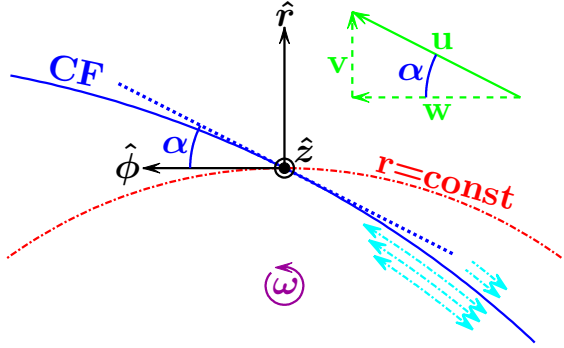}}}
	\caption{
		\label{fig:Defs}
		Illustration of the two-phase spiral flow model in the equatorial plane.
		The tangent (dotted line) to the CF (solid curve) forms an angle $\alpha$ with $\unit{\phi}$, \ie with the circle $r=\constant$ (dot-dashed).
		The plasma must flow parallel to the CF.
		The planar velocity $\vect{u}$ (thick green arrow; parallel to the CF at the point analyzed) is decomposed into radial ($\myv$) and azimuthal ($\myw$) components (dashed green arrows). The flow above the CF is a slow inflow (dot-dashed cyan arrows). The flow below the CF is a fast \citep{Keshet10} inflow or outflow (dot-dashed cyan double headed arrows).
		The pattern rotates (circular arrow) about the $z$-axis, in the $\pm\unit{z}$ sense for $\omega\gtrless 0$.
	}
\end{figure}

Motivated by the extended spiral patterns recently observed, we generalize the core spiral flow model of \citetalias{Keshet12} to the entire GA.
The model focused on the two plasma phases flowing above and below the CF.
Below the CF, a cold, fast flow is inferred from hydrostatic non-equilibrium measurements in many galaxy clusters
\citep{Markevitch01, Keshet10}; this could be either an inflow or an outflow. Above the CF, the hot phase must be a slow inflow \citep{Keshet10};
see the illustration in Figure \ref{fig:Defs}.

As the spiral has no natural scale, the model is constructed using power-law radial profiles. The interaction between the slow and fast flow components is mediated by the $\partial_\phi P \propto P$ azimuthal pressure gradient near the CF. Generalizing the model to include the gradual variation in the CF contrast $q$ over a wide radial span, such that $\partial_\phi P \propto qP$ in the CF vicinity, yields a more realistic core structure. Surprisingly, it also reproduces the universal profile of Equations~(\ref{eq:linear_entropy})--(\ref{eq:K fit_per_GA}) throughout the GA, independent of the gravitating mass or pressure profiles. This model, focusing on the CF vicinity, is independent of the arguments made in Section~\ref{sec:spiral origin} based on the radially declining azimuthal gradients far from the CF.

Assuming a steady state in a frame rotating at some fixed angular frequency $\omega$, the flow is governed by the time independent continuity equation,
\begin{equation}
\label{eq:continuity}
\grad\cdot \left(\rho \mathbf{U}\right)=0\, ,
\end{equation}
the momentum equation,
\begin{equation}
\label{eq:momentum}
\mathbf{U}\cdot\grad\mathbf{U}=-\rho^{-1}\grad P + \mathbf{g} - 2\mathbf{\Omega}\times\mathbf{U}-\mathbf{\Omega}\times(\mathbf{\Omega}\times\mathbf{r})\, ,
\end{equation}
and the energy equation,
\begin{equation}
\label{eq:energy}
\frac{\kB\myzeta n_e}{\Gamma-1}\left(\mathbf{U}\cdot\grad T+T\grad\cdot \mathbf{U}\right)=\grad\cdot\left(\kappa\grad T\right)-n_e^2\Lambda\left(T\right)\, .
\end{equation}
Here, vectors such as the position vector $\mathbf{r} = \{r,\phi,z\}$ and the velocity vector $\mathbf{U} = \{\myv,\myw,U_z\}$ are written in cylindrical coordinates, $\mathbf{\Omega}=\omega \mathbf{\hat{z}}$ is the angular velocity vector, $\mathbf{g}$ is the gravitational acceleration (dominated by dark matter), and $\Gamma=5/3$. For simplicity, consider simple power-law profiles for $n_e$ and $T$.

\subsection{Spiral Flow Equations}
Consider first the flow confined to the equatorial plane.
We use the notations of \citetalias{Keshet12} (see Figure \ref{fig:Defs}), but extend the analysis to the whole GA. Due to the presence of a CF, the flow (Equations \eqref{eq:continuity}-\eqref{eq:energy}) in its vicinity can be cast in a simple form (\citepalias{Keshet12}; continuity),
\begin{equation}
\label{eq:cont-lambda}
\lambda_\rho = -\frac{r}{\myv}\grad\cdot\mathbf{U}=-\left(1+\mypw+\tilde{\lambda}_z\right)\, ,
\end{equation}
momentum conservation along $\phi$,
\begin{equation}
\frac{d}{dr}\left(\frac{\myw^2}{2}\right)+\frac{\myw^2}{r}=\frac{a_\phi}{\gamma}-2\omega \myw\, ,
\end{equation}
momentum conservation along $r$,
\begin{equation}
\frac{d}{dr}\left(\frac{\myv^2}{2}\right)-\frac{\myw^2}{r}=a_r+\omega^2 r + 2\omega \myw\, ,
\end{equation}
and energy conservation,
\begin{equation}
\label{eq:energy-lambda}
\frac{\myv\rho T}{r}\left(1+\mypw+\frac{\lambda_T}{1-\Gamma}+\tilde{\lambda}_z\right)=-n_e^2\Lambda(T)\, ,
\end{equation}
where full derivatives are taken along the flow,
\begin{equation}
\frac{d}{dr}=\frac{\partial}{\partial r}+\frac{1}{\gamma r}\frac{\partial}{\partial \phi}\, .
\end{equation}
Here, logarithmic derivatives $\lambda_A \equiv d \mathrm{ln} |A|/d \mathrm{ln} \,r$ are used (to represent an arbitrary quantity $A$) and $\mathbf{a} = \{a_r,a_\phi,a_z\}= - \rho^{-1}\grad P + \mathbf{g} $ is the inertial acceleration.
Thermal conduction is neglected in this part of the flow.

These equations were derived assuming that an equatorial plane can be found, perpendicular to the $z$ symmetry axis, in which all streamlines are approximately confined to the plane. However, they hold throughout the volume if $\mathbf{U}$ and the CF surface are perpendicular when projected onto the $z$-$r$ plane; see \citetalias{Keshet12}. The radius of curvature perpendicular to the plane, $R_\theta= −\left[1+r'(z)^2\right]^{3/2}/r''(z)$, enters the equations through the relation
\begin{equation}
\tilde{\lambda}_z \equiv \frac{r}{\myv}\frac{\partial U_z}{\partial z} = \frac{r}{R_\theta}\, .
\end{equation}

\subsection{The Fast, Cold Component}
An important clue to the nature of core spirals is the presence of fast, nearly sonic flows, found beneath well-observed CFs in the core \citep{Keshet10}. This flow leaves the system before it significantly cools, and so it is nearly adiabatic. Equations~\eqref{eq:cont-lambda} and \eqref{eq:energy-lambda} then fix its scaling \citepalias{Keshet12},
\begin{equation}
\lambda_{\rho}^f=\frac{\lambda}{\Gamma}\:;\quad
\lambda_{T}^f=\frac{\Gamma-1}{\Gamma}\lambda\:;\quad
\mbox{and}\quad
\frac{r}{R_\theta}=-\frac{\lambda}{\Gamma}\:,
\end{equation}
as a function of the logarithmic pressure derivative, $\lambda\equiv\lambda_P$,
and
\begin{equation}
\label{eq:lambda_w}
\mypwf=-1\:.
\end{equation}
Here, subscripts $f$ $(s)$ denote the fast (slow) component.

\subsection{The Slow, Hot Component}

Although azimuthal pressure gradients are small, they must be nonzero in order to allow for a slow component \citepalias{Keshet12}, and are indeed directly observed (Figure~\ref{fig:PerseusPressure}). To first order, one may assume \citepalias{Keshet12} that $\partial_\phi P \propto P$. However, in order to generalize the analysis to the whole GA, here we must take into account the variation in the CF contrast $q$ with radius.
We therefore assume that the azimuthal acceleration $a_\phi$ of the slow component due to the pressure gradient is proportional to $q$,
\begin{equation}
\label{eq:anstaz}
2\omega \myv \simeq a_\phi \simeq -\frac{q}{r\rho}\partial_\phi P\:.
\end{equation}

This naturally leads to the slow velocity scaling
\begin{equation}
\label{eq:momentum-phi-lambda}
\mypvs=\lambda_q+\lambda-1-\lambda_\rho^s\:,
\end{equation}
where the isobaric CF contrast follows
\begin{equation}
\label{eq:lambda_q}
\lambda_q=\lambda_T^s-\lambda_T^f=\lambda_\rho^f-\lambda_\rho^s\:.
\end{equation}
For the approximate $\Lambda \propto T^{1/2}$ cooling function, the energy equation \eqref{eq:energy-lambda} for the slow, cooling flow becomes
\begin{equation}
\mypvs+\frac{1}{2}\lambda_T^s-1\simeq\lambda_\rho^s\:.
\end{equation}

Combining these with with Equation \eqref{eq:cont-lambda} gives the scaling of the slow component,
\begin{equation}
\lambda_\rho^s\simeq\frac{1}{7}\left[-4+\left(3+\frac{2}{\Gamma}\right)\lambda\right]=-\frac{4}{7}+\frac{3}{5}\lambda\:,
\end{equation}
\begin{equation}
\label{eq:slow T}
\lambda_T^s\simeq\frac{-2\lambda+4\Gamma(1+\lambda)}{7\Gamma} =\frac{4}{7}+\frac{2}{5}\lambda\:,
\end{equation}
\begin{equation}
\label{eq:s_v}
\mypvs\simeq\frac{\Gamma+(3+\Gamma)\lambda}{7\Gamma} =\frac{1}{7}+\frac{2}{5}\lambda\:,
\end{equation}
and
\begin{equation}
\label{eq:s_w}
\mypws\simeq\frac{-3\Gamma+(5-3\Gamma)\lambda}{7\Gamma} =-\frac{3}{7}\:,
\end{equation}
the latter being independent of $\lambda$ for $\Gamma=5/3$.

For the entropy of the slow phase, we finally obtain
\begin{equation} \label{eq:SpiralModel_K}
\lambda_K^s \simeq \frac{4\Gamma+(5-3\Gamma)\lambda}{7} =\frac{20}{21}\simeq 1\:,
\end{equation}
flatter than found by assuming $\partial_\phi P\propto P$ in \citetalias{Keshet12}.
As the slow component dominates the flow by mass, and strongly dominates it by volume, we approximate the overall entropy profile as $\lambda_K\simeq \lambda_K^s\simeq 1$.

The entropy power-law index in Equation~(\ref{eq:SpiralModel_K}) is very similar to the observed, universal profile in Equations~\eqref{eq:K fit} and \eqref{eq:K fit_per_GA}.
Over the wide radial range considered, the pressure profile cannot be approximated as a pure power law, so $\lambda$ is not constant.
In contrast to most other properties in the model, the entropy profile turns out to be independent of this $\lambda$ (for $\Gamma=5/3$).
Hence, the entropy profile is robustly found to be a pure power law, independent of the underlying gravitating mass.

Equation \eqref{eq:SpiralModel_K} was derived for a $\Lambda(T)\propto T^{1/2}$ cooling function.
We find that $\lambda_K$ changes very little, by $[-2,+6]\%$, for more realistic cooling functions; at large radii it becomes even closer to unity.

\subsection{Properties of the generalized model}

It is interesting to examine the other properties of the generalized model. First, note that as $\rho^{-1}\partial_r P$ is fixed by the gravitational potential (to zeroth order, using hydrostatic equilibrium), and the entropy profile of the model matches the linear profile observed, the radial profiles of both the density and the temperature are guaranteed to match observations, both inside and outside of the core.

Next, consider the spiral structure. Using Equations \eqref{eq:s_v} and \eqref{eq:s_w}, the spiral pattern is given by
\begin{equation}
\lambda_\gamma=\mypvs-\mypws\simeq\frac{-2\lambda+4\Gamma(1+\lambda)}{7\Gamma}=\frac{4}{7}+\frac{2}{5}\lambda\:,
\end{equation}
which, using Equation \eqref{eq:lambda_w}, determines the fast radial velocity scaling,
\begin{equation}
\label{eq:lambda_v}
\mypvf=\lambda_\gamma+\mypwf\simeq
\frac{1}{7}\left[-3+\left(4-\frac{2}{\Gamma}\right)\lambda\right]=-\frac{3}{7}+\frac{2}{5}\lambda\:.
\end{equation}

Now, using Equations~\eqref{eq:s_v} and \eqref{eq:lambda_v}, the radial dependence of the shear across the spiral CF is found to follow
\begin{equation}
\mypvf-\mypvs=\frac{-4 + \left(3 - 5\Gamma\right) \lambda}{7} = -\frac{4}{7}\:,
\end{equation}
and the CF contrast (Equation \ref{eq:lambda_q}) to scale as
\begin{equation}
\lambda_q=\frac{4 - \left(3 - 5\Gamma\right) \lambda}{7} = \frac{4}{7}\:,
\end{equation}
more gradual than the $-4/5$ and $+4/5$ indices of \citetalias{Keshet12}.
A summary of the scaling in the model is given in Table~\ref{table:value}.

\bgroup
\def\arraystretch{1.75}
 \begin{table*}
 \caption{\label{table:value} Scaling of the two-component spiral model, for $\Gamma=5/3$
 }
 \begin{center}
 \begin{tabular}{l|c|c}
 \hline
 Property $A$ & \multicolumn{2}{c}{Logarithmic derivative $\lambda_A=d\log|A|/d\log(r)$}   \\
 \hline
 Pressure $P$ & \multicolumn{2}{c}{$\lambda\equiv \lambda_P$} \\
 Spiral slope $\gamma$ & \multicolumn{2}{c}{$\frac{-2\lambda+4\Gamma(1+\lambda)}{7\Gamma}=\frac{4}{7}+\frac{2}{5}\lambda$} \\
  Curvature & \multicolumn{2}{c}{$\frac{r}{R_\theta} = -\frac{\lambda}{\Gamma}$}
  \\
 \hline
  & Fast phase & Slow phase \\
 \hline
 Density $\rho$ & $\frac{\lambda}{\Gamma}=\frac{3\lambda}{5}$    & $\frac{-4\Gamma+\left(3\Gamma+2\right)\lambda}{7\Gamma}=-\frac{4}{7}+\frac{3}{5}\lambda$ \\
 Temperature $T$ & $\frac{\Gamma-1}{\Gamma}\lambda=\frac{2\lambda}{5}$     & $\frac{-2\lambda+4\Gamma(1+\lambda)}{7\Gamma} =\frac{4}{7}+\frac{2}{5}\lambda$ \\
 Radial velocity $\myv$ & $\frac{-3\Gamma+\left(4\Gamma-2\right)\lambda}{7\Gamma}=-\frac{3}{7}+\frac{2}{5}\lambda$
 & $\frac{\Gamma+(3+\Gamma)\lambda}{7\Gamma} =\frac{1}{7}+\frac{2}{5}\lambda$ \\
 Tangential velocity $\myw$ & $-1$    & $-\frac{3\Gamma-(5-3\Gamma)\lambda}{7\Gamma} =-\frac{3}{7}$ \\
 Entropy $K$ & $0$ & $\frac{4\Gamma+\left(5-3\Gamma\right)\lambda}{7} =\frac{20}{21}$ \\
 \hline
 \end{tabular}

 \end{center}
 \end{table*}

In addition to the more realistic thermal profiles of the generalized model, it also gives a more feasible dynamical structure. First, the radial decline in the shear and in the velocity of the fast phase, and the rise in the CF contrast are more gradual here, allowing for a more extended structure.
The spiral pattern now scales as the temperature of the slow component, which dominates the flow, $\lambda_\gamma=\lambda_T^s\simeq \lambda_T$. This corresponds to a purely logarithmic spiral at the edge of the core, becoming slightly Archimedean (hyperbolic) outside (inside) the core.
The perpendicular structure $\tilde{\lambda}_z$ is unchanged by our generalization, giving a prolate (closer to cylindrical) structure inside the core.
At the larger radii considered here, the 3D structure becomes nearly spherical at the edge of the core, and oblate near the virial radius.

Although the contrast and shear change more slowly in the generalized model, they do vary significantly over the larger radial span considered.
This may lead to large shear and contrast, in excess of observational constraints, if a spiral pattern is assumed to continuously extend all the way from the inner core ($\sim10\kpc$) to the GA periphery ($\sim2\Mpc$), as even a modest ($q=1.1$, say) CF at $20\kpc$ would imply an unrealistic, $q\gtrsim 10$ beyond $1\Mpc$. This would contradict the $q\sim1.5$ contrasts observed at large radii \citep{Rossetti13, Urban14, Walker14}.

An alternative is that instead of one continuous spiral, the CF is actually intermittent, composed of a series of spiral sections separated by small gaps. This possibility may be supported by the intermittent spiral patterns and CF gaps, often observed on both small \cite[\eg][]{Churazov03} and large scales \citep{Urban14,Walker14}.

\section{Discussion}
\label{sec:Discussion}

We study two apparently independent, universal properties of GAs: a simple power-law entropy profile extending out to the virial radius, and a ubiquitous underlying spiral pattern, apparently reaching out to the virial radius as well.
We show that the former is mass-independent over a wide mass range, and argue that it reflects a robust dynamical mechanism, regulating the thermal structure of the GA; models have been unable to naturally reproduce it until now.
The latter reflects a long-lived, extended spiral flow, present in many, if not all, GAs.
Furthermore, speculating that the two phenomena are related, we demonstrate that an extended spiral flow can naturally explain the universal thermal profile, under simple (yet somewhat ad hoc) assumptions regarding its structure.
Interestingly, three independent physical arguments lead to approximately the same thermal profile, supporting these assumptions.

To quantify the thermal profile, we analyze 16 deprojected clusters and 12 deprojected groups from the literature (see Table \ref{table:my data table}), and find a good, nearly linear entropy $K(r)$ fit (see Equations \eqref{eq:linear_entropy}--\eqref{eq:K fit_per_GA}, and Figures \ref{fig:entropy scaling}--\ref{fig:K100 by cluster}).
The power-law index, $\lambda_K = 0.96\pm0.01$, is consistent with the mean $\lambda_K\sim(0.9\till 1.1)$ of previous studies, which used slightly different methods (see Table \ref{table:entropy fits}).
Our tight fit is due to the wide mass range, the minimization of projection effects, and an $r>10\kpc$ selection of fit data.

A different parameterization, suggested by \cite{McCourt12}, involves the ratio $\tau$ between the cooling and free-fall times.
Our data is similarly well-fit by the above $K\propto r^{\lambda_K}$, and by the relation (see Equation~\ref{eq:simple fit} and Figure \ref{fig:universal scaling}) $\tau\propto r^{0.72\pm0.01}$.
We are unable to determine which of these two profiles
is more realistic, as they differ only by a $T^{1/3}$ factor, whereas the temperature range spanned by our data is narrow and the uncertainties are substantial.
Both profiles slightly flatten for $r\lesssim 10\kpc$.
This is where $\tau\sim 1$, suggesting that $\tau$ may play a more fundamental role than $K$ in defining the thermal profile.

Previous suggestions to scale the radius by $\myR_{500}$, the entropy by $T^{0.65}$, or both, only diminish the correlation; the $\myR_{500}$ scaling effectively splits the GAs by mass.
We find no difference between GAs with versus those without H$\alpha$ filaments, nor any $\tau(r)$ profiles showing a minimum; this differs from previous reports, but is consistent with the absence of an `entropy floor' \citep[\eg][]{Panagoulia14}.

We find that the giant elliptical galaxy M49, which resides in the outskirts of the Virgo cluster and harbors an AGN, approximately follows the same universal thermal profile (see Figures \ref{fig:entropy scaling} and \ref{fig:universal scaling}). It would be interesting to see if the universal profile characterizes all AGN harboring systems, and thus more generally applies not only for GAs.

The universal thermal profile may seem surprising, as it appears to be oblivious to the temperature peak at the edge of the core and to the virial shock at the edge of the GA, and is robust to the presence of the ongoing cooling, merger, and AGN activity.
Furthermore, the profile (in either of the two idealized forms we derive, Equations~\eqref{eq:conduction scale} or \eqref{eq:entropy scale}) is not naturally obtained from the flow equations without introducing some large number or a dimensional constant, even if conduction and cooling are taken into account.
This suggests some non-local mechanism.

The entropy profile is usually attributed to the self-similar evolution of primordial gas crossing the virial shock.
However, we find that this model, without baryonic effects, does not provide a good description to the data, for the following reasons:
\emph{(i)} the observed profile shows no break as one crosses into the core, where radiative cooling becomes significant and the self-similarity is expected to break;
\emph{(ii)} the self-similar model predicts a $K_{100}\propto M^{0.26}$ correlation between the entropy and the GA mass, which is inconsistent with our results (see Figure~\ref{fig:K100 by cluster}) by more than $8\sigma$;
\emph{(iii)} the normalization of the entropy in the self-similar model is inconsistent with our results (at significance levels ranging between $1\sigma$ and $20\sigma$ for different clusters, and $>6.5\sigma$ on average; see Figure~\ref{fig:K100 by cluster});
\emph{(iv)} The self-similar model predicts a radial power law $\lambda_K\approx 1.2$ \citepalias{Voit05}, which is inconsistent with our results at $>8\sigma$ for the full $r>10\kpc$ sample, or $>4\sigma$ when limited to radii outside the core ($0.3<r/\myR_{500}<1.5$; see Figure~\ref{fig:entropy scaling}).

Alternative mechanisms, such as non-standard heat conduction or turbulent heating, require fine tuning.
We therefore argue that a robust, dynamical, non-local mechanism is needed to sustain the universal profile.

A natural candidate for such a dynamical mechanism involves the spiral flow, as it too is ubiquitous, spans the same radial range, and similarly overlaps the CC where radiative cooling plays an important role.
This suggests that the universal profile and the spiral flow can both be regarded as manifestations of the cooling problem.
To examine if a spiral flow can regulate the structure of the GA and enforce the universal thermal profile, some assumption must be made regarding the azimuthal gradients, in particular those of pressure (\eg Figure \ref{fig:PerseusPressure}) or temperature.

The simplest assumption involving a new length scale, motivated by the slight Archimedean nature of CC spirals \citepalias{Keshet12} and by the necessary presence of a convective layer (Appendix \ref{sec:Schwarzschild}), is that in some part of the flow, azimuthal gradients diminish with $r$ as $\partial_\phi\log A\sim L/(2\pi r)$, where $L$ is an a priori unknown length scale. A balance between cooling and azimuthal heat conduction thus recovers the $K(r)\propto r$ profile (see Equations~\eqref{eq:entropy scale} and \eqref{eq:Spiral_K}), while independently requiring a balance between cooling and radial heat advection reproduces the $\tau\propto r^{3/4}$ relation (see Equations~\eqref{eq:conduction scale} and \eqref{eq:Spiral_tRatio}).
Interestingly, although these two arguments are of a very different physical nature, both give similar $L\sim 1\till10\kpc$ scales.
Moreover, this range of $L$ turns out to be quite natural, signaling the regime where $\tau$ approaches unity and the thermal profile flattens, so no artificial scale is introduced.
Indeed, this is also the typical radius at the base of the spiral, where radial and azimuthal gradients are expected to become comparable.

These conclusions confirm the dynamical origin of the universal profile, support the ad hoc azimuthal gradient assumption, and provide hints for more sophisticated spiral flow modeling.
The universal thermal profile, in particular its normalization, shows no correlation with GA mass, as demonstrated in Figures.~\ref{fig:entropy by cluster} and \ref{fig:K100 by cluster}.
This strongly constrains the mechanism behind the thermal profile.
In the simple spiral flow modeling of section \ref{sec:spiral origin}, this corresponds to the length scale $L$ being a universal, mass-independent constant.

A third independent way to see how a spiral flow could regulate the universal profile is to analyze the flow along the CFs. To do so we generalize the core spiral flow model of \citetalias{Keshet12} to the entire GA.
Here, the flow is approximated as a combination of fast and slow components, interacting through the azimuthal pressure gradients in the CF vicinity.
To extend the model over a large radial span, we incorporate the CF contrast $q$ to obtain $\partial_\phi P \sim qP$.
This expression yields more realistic thermal profiles within the core, and a more feasible dynamical structure than found in \citetalias{Keshet12}.
Surprisingly, it also reproduces an approximately linear $K(r)$ profile throughout the GA (see Equation \ref{eq:SpiralModel_K}), consistent with observations, and independent of the (non-power-law) radial pressure profile.
Consequently, all the observed thermal profiles are reproduced, regardless of the underlying mass distribution, both in and beyond the core, out to the virial radius.

A spiral flow may alleviate the cooling problem (\citetalias{Keshet12}), because it can (i) advect heat inward from the periphery or outward from the AGN; (ii) distribute energy evenly in space and time; (iii) provide a feedback loop by regulating the flow rate; and (iv) operate with little to no mass accretion.
We strengthen this model by showing how a careful balance between cooling, radial heat advection, and azimuthal heat conduction consistently yields the universal profile under plausible assumptions.

The generalized two-component spiral model provides additional information regarding the thermal and dynamical structure of GAs (see Table \ref{table:value}).
Most importantly, such models yield a mild temperature drop toward the GA center, avoiding the cooling problem (\citetalias{Keshet12}).
With increasing $r$, the CF manifold is predicted to gradually transition from mildly prolate to oblate, and from a slightly hyperbolic to a mildly Archimedean spiral.
The CFs in the model show an increasing $q\propto r^{4/7}$ contrast, and a decreasing $\sim r^{-4/7}$ shear.
Although this allows for more extended CFs than in \citetalias{Keshet12}, such CFs cannot continuously extend from the GA center out to its outskirts, suggesting some pattern intermittency, in accord with observations \cite[\eg][]{Churazov03,Urban14,Walker14}.

A central open question is the mechanism driving the spiral flow and depositing the necessary angular momentum; see discussion in \citetalias{Keshet12}. Our results do not distinguish between the two previous suggestions for the mechanism driving the spiral - either mergers or strong AGN activity.
Our assumptions regarding the azimuthal gradients are somewhat ad hoc, as one could have conjectured more complicated azimuthal profiles such as $\partial_\phi P \propto (L/r)^\beta P$ with some $\beta\neq 1$, or $\partial_\phi P \propto (q-1)P$. In general, however, these would have led to non-power-law scalings, or to inconsistencies.
A more sophisticated and in part numerical modeling of spiral flows is needed in order to quantify these gradients and resolve the 3D structure of the flow.

Spiral flows stabilized by shear-amplified magnetic fields, featuring a fast flow under the CF, are already found in magnetohydrodynamic (MHD) simulations of single clusters undergoing a merger \citep{ZuHone11}.
Spiral flows were also produced in simulations incorporating radiative cooling \citep[severely limiting their lifetime;][]{ZuHone10} and different forms of viscosity \citep{Roediger13} and heat conduction \citep{ZuHone13}.
To produce a physically viable model for large-scale spiral flows, simulations would need to combine the above elements in a self-consistent fashion.
Such simulations would need to resolve the spiral structure both at its base ($\lesssim 1\kpc$) scale, and its outer (virial) scale.
The observational evidence outlined above suggests that additional physical processes, notably the virial shock and AGN feedback, \eg in the form of radio bubbles, are important ingredients of the spiral flow and should be incorporated into simulations.
A successful simulation should show a long-lived core, and reproduce the observed radial scalings, in particular the universal profile and the structure of the 3D CF surface gradually uncovered by observations.

Such simulations, as well as future observations, will directly test our results.
In particular, they could either refine or rule out our spiral flow suggestion for the dynamical mechanism regulating the universal profile.
For example, the presence of a fast flow below the CF, which is based on present observations and simulations, is a key assumption of the spiral flow model; it will be critically tested with far greater accuracy by future observations and simulations.
A realistic spiral flow simulation, incorporating the physical processes outlined above, would directly test our approximation of two-phases composing the spiral, and gauge the simplified power-law scalings we derive.
Simulations and observations would also test additional predictions of the model, such as the presence of a convective zone, a CF intermittency, and entrained radio bubbles \citep{Keshet12}.
The role of the spiral flow in regulating the universal profile will be ruled out if this profile is observed in the clear absence of any underlying spiral pattern, or if simulations conclusively show that spiral flows are non-universal, \eg greatly depending upon initial conditions.

Future deprojected profiles of GAs, in particular at the low and high temperature ends, should distinguish between the two forms of the universal profile (Equations~\ref{eq:K fit} versus \ref{eq:simple fit}).
Deep observations of GAs can place strong constraints on the spiral pattern, on the structure of the CF surface, and on the flow itself through, for example, line-of-sight velocity measurements.
These constraints would allow for a better modeling of the underlying spiral flow, and would directly test the azimuthal gradients postulated above, as well as the flow and CF properties they imply.
Our results suggest that the diffuse gas survives down to scales where $\tau\sim 1$, smaller than thought so far, where the strong competition between gravity and cooling is likely to have interesting consequences.

The identification of spiral structures in galaxy clusters is challenging even at small radii, due to its subtlety and to
resolution and projection effects. Identifying spiral structures at large radii is even more difficult, due to the lower $\SB$. Therefore, the recently identified large-scale spiral structures in three clusters -- Perseus \citep{Simionescu12}, A2142 \citep{Rossetti13}, and RXJ2014 \citep{Walker14} -- and observational hints of such structures in several more deep observations \cite[\eg A496, A3558, A85, A2052, and A2199; see][]{Rossetti14} suggest that this phenomenon is ubiquitous. In our spiral model for the universal profile, spiral structures are inevitable at large radii. Deeper observations will test this prediction.

\acknowledgements
This research has received funding from the European Union Seventh Framework Programme (FP7/2007-2013) under grant agreement n\textordmasculine~293975, from an IAEC-UPBC joint research foundation grant, and from an ISF-UGC grant.

\appendix

\section{Testing a possible entropy flattening at the outskirts}
\label{sec:flat}

To further test any possible flattening of the entropy in the outskirts of clusters \citep[\eg the \emph{Suzaku} analysis in][]{Walker12}, Figure \ref{fig:flat} shows the entropy in our sample using the method of \cite{Walker12}; namely, scaling the entropy to its value at $0.5\myR_{500}$ and to a linear radial profile, and scaling the radius to $\myR_{500}$.
Unlike \cite{Walker12}, we see no significant flattening in the entropy at cluster outskirts, in accordance with stacked \emph{ROSAT} \citep{Eckert12} and \emph{Chandra} \citep{MorandiEtAl15} observations.
This has been discussed, based on the data used here, in \cite{Eckert12}.

\begin{figure}
	\centerline{\epsfxsize=13.4cm \epsfbox{\myfig{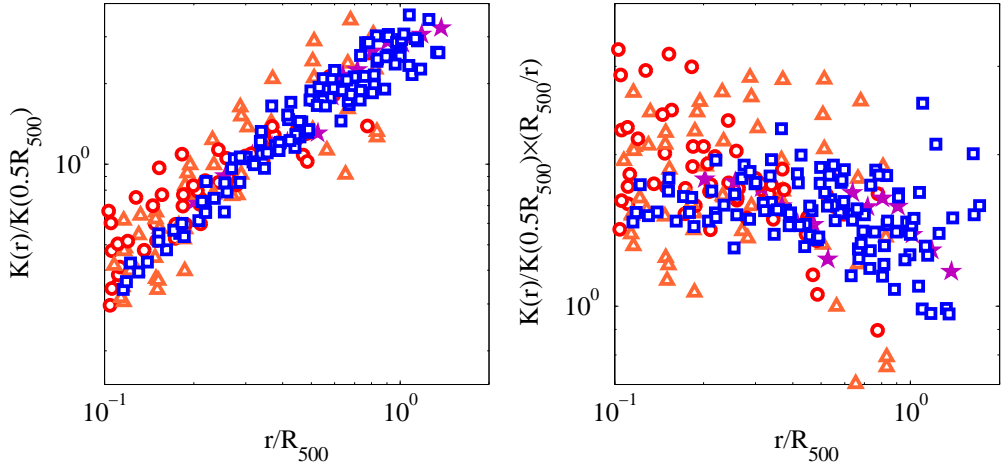}}}
	\caption{\label{fig:flat} Left: the entropy (scaled to the entropy at $0.5\myR_{500}$) as a function of the scaled radius.
Right: radially normalized entropy, $K(r)/r$ (scaled to $0.5\myR_{500}$), plotted to enhance any deviation from a power-law behavior. The symbols are the same as in Figure \ref{fig:entropy scaling}.
	}
\end{figure}

\section{Generalized Schwarzschild criterion}
\label{sec:Schwarzschild}
We extend the Schwarzschild criterion \citep{Schwarzschild58,Lebovitz65} for convective stability, $\partial_rK>0$, to an anisotropic medium harboring a spiral flow, implying that a mixing layer must form in a realistic spiral flow. Such flows show CFs with an entropy contrast $q_K=K_o/K_i$, where subscript $i$ ($o$) represents the region just inside (outside) the CF. Therefore, the azimuthal entropy derivative must at least somewhere satisfy $\partial_{\phi}K_i\geq(q_K-1)K_i /2\pi$. At that point, convective stability against perturbations along both $r$ and $\phi$ would require
\begin{equation}
\label{eq:stability crierion}
\lambda_K\simeq\lambda_{K_i}= \frac{r}{K_i}\left(\frac{\partial K_i}{\partial r}+\frac{1}{\gamma r}\frac{\partial K_i}{\partial\phi}\right)>\frac{1+\gamma}{\gamma K_i}\frac{\partial K_i}{\partial\phi}\geq\frac{(q_K-1)(1+\gamma)}{2\pi\gamma} \, ,
\end{equation}
where the first approximate equality assumes a self-similar spiral structure, so $\lambda_K$ can be measured just inside the CF, and in the first inequality we require that azimuthal perturbations do not lead to larger radial displacements; the (weaker) Schwarzschild criterion would merely remove the $\gamma$ term in the numerators. The entropy profile approximately follows $\lambda_K\simeq1$, the spiral pattern roughly shows $\gamma\simeq0.2$ \citepalias{Keshet12}, and the temperature contrast $q$ yields $q_K\simeq q^{5/3}\simeq3$. Hence, Equation \eqref{eq:stability crierion} breaks down somewhere, and a convective layer is expected to form.

Equation \eqref{eq:stability crierion} indicates that the azimuthal derivative of the entropy cannot be constant, as the entire medium will then be convectively unstable. Hence, the derivative must diminish across a large sector, as assumed in Section~\ref{sec:spiral origin}.

\def\apj{ApJ}                 
\def\apjl{ApJL}                
\def\aap{A\&A}                
\def\mnras{MNRAS}             
\def\pasj{PASJ}               
\def\physrep{Phys.~Rep.}   
\def\jcap{J. Cosmology Astropart. Phys.}
\def\araa{ARA\&A}             

\vspace{1cm}


\label{lastpage}

\end{document}